\documentclass[twocolumn,english,prl,amsmath,superscriptaddress,showpacs]{revtex4}
\usepackage{ae,aecompl}
\usepackage[T1]{fontenc}
\usepackage[latin9]{inputenc}
\usepackage{units}
\usepackage{textcomp}
\usepackage{amsmath}
\usepackage{graphicx}
\usepackage{amssymb}

\DeclareRobustCommand{\greektext}{%
  \fontencoding{LGR}\selectfont\def\encodingdefault{LGR}}
\DeclareRobustCommand{\textgreek}[1]{\leavevmode{\greektext #1}}
\DeclareFontEncoding{LGR}{}{}

\usepackage{babel}

\begin{document}

\title{Spin relaxation in sub-monolayer and monolayer InAs structures grown
in GaAs matrix }

\author{Chunlei Yang}

\affiliation{Department of Physics, The University of Hong Kong, Hong Kong, China}

\affiliation{School of Physics and Engineering, Sun Yat-Sen University, Guangzhou,
China}

\author{Xiaodong Cui}

\email[Email:]{xdcui@hku.hk}

\affiliation{Department of Physics, The University of Hong Kong, Hong Kong, China}

\author{Shun-Qing Shen}

\affiliation{Department of Physics, The University of Hong Kong, Hong Kong, China}

\author{Zhongying Xu}

\affiliation{State Key Laboratory for superlattices and microstructures, Institute
of Semiconductors, Chinese academy of Sciences, Beijing, China }

\author{Weikun Ge}

\affiliation{School of Physics and Engineering, Sun Yat-Sen University, Guangzhou,
China}

\date{\today}

\pacs{73.21.Fg, 72.25.Rb, 78.47.jc, 78.47.jc}
\begin{abstract}
Electron spin dynamics in InAs/GaAs heterostructures consisting of
a single layer of InAs (1/3$\sim$1 monolayer) embeded in (001) and
(311)A GaAs matrix was studied by means of time-resolved Kerr rotation
spectroscopy. The spin relaxation time of the sub-monolayer InAs samples
is significantly enhanced, compared with that of the monolayer InAs
sample. We attributed the slowing of the spin relaxation to dimensionally
constrained D\textquoteright{}yakonov-Perel\textquoteright{} mechanism
in the motional narrowing regime. The electron spin relaxation time
and the effective $g$-factor in sub-monolayer samples were found
to be strongly dependent on the photon-generated carrier density.
The contribution from both D\textquoteright{}yakonov-Perel\textquoteright{}
mechanism and Bir-Aronov-Pikus mechanism were discussed to interpret
the temperature dependence of spin decoherence at various carrier
densities.
\end{abstract}
\maketitle

\section{Introduction }

One of pioneering approaches towards prospective spintronic devices
is to manipulate electron spins by utilizing spin-orbit coupling in
non-magnetic semiconductors, particularly in low-dimensional III-VI
semiconductor heterostructures (quantum wells, wires, and dots) owing
to their great flexibility in manipulating spin properties of the
electronic states.\cite{Awschalom1} Unlike electron charge, electron
spin is not conserved and generally relaxes to un-polarized states
in solids. There exist competing spin relaxation channels: spin-flip
through electron-impurity scattering, known as Elliott and Yafet (EY)
mechanism,\cite{EY} spin-flip through electron-hole exchange scattering
known as Bir\textendash{}Aranov\textendash{}Pikus (BAP) mechanism,\cite{BAP}
and spin-flip through spin-orbital coupling known as D\textquoteright{}yakonov\textendash{}Perel\textquoteright{}
(DP) mechanism.\cite{DP} The relative importance of these mechanisms
is strongly dependent on semiconductor structure, temperature and
carrier concentration. 

InAs/GaAs heterostructures as one of the potential spintronic building
blocks have received intensive attention for a couple of decades.
Since the quantum islands formed in InAs submonolayers display a narrow
size distribution as revealed by their sharp PL spectrum ($<$2 meV),
it is of advantage to utilize spin states of submonolayer structures
for the sake of almost unified electronic states which are hard to
achieve for self-assembled quantum dots due to wide size distribution.
Nevertheless, there are few experiments addressing the spin dynamics
in ultrathin InAs layers.

In this paper, we study the spin relaxation in InAs/GaAs heterostructures
consisting of a single layer of InAs with respectively effective thickness
of 1/3, 1/2 and 1 monolayer by time resolved Kerr rotation spectroscopy.
The spin relaxation time is found to be increased with decreased InAs
coverage. This behavior could be attributed to lateral restriction
under the DP mechanism.

\section{Experiment}

Ultrathin InAs layer is a kind of InAs/GaAs heterostructure in which
a single InAs layer is sandwiched in a GaAs matrix. The samples were
grown by elemental source MBE on (001) and (311)A-oriented GaAs semi-insulating
substrates. The structures consist of a GaAs buffer layer, cladding
layers and an InAs layer. Each cladding layer is composed of 40 period
GaAs/Al$_{\text{0.4}}$Ga$_{\text{0.6}}$As superlattices (SL) and
a GaAs layer. The use of GaAs/AlGaAs SL improves sample quality by
preventing surface recombination and trapping defects. The single
InAs layer with various effective thicknesses is sandwiched between
the cladding layers. The details of the growth could be found elsewhere.\cite{MBE_method}
To study the lateral size effect on the spin relaxation process, three
samples with 1/3, 1/2 and 1 monolayer (ML) InAs, respectively, on
(001) substrates and one with 1 ML InAs on (311)A substrate were prepared.
The 1ML InAs layer on (001) surface is assumed to be an ideal two
dimensional (2D) system. Instead, due to the intrinsic surface corrugation,\cite{corrugation}
the 1ML InAs layer on (311)A surface forms wire or disk like micro-structures
on the GaAs steps and facets. While for the 1/2 ML and 1/3 ML samples
on (001) surface, InAs is found to be organized as wire-like or disk-like
islands elongated along {[}$01\overline{1}$] direction with lateral
size in the range of tens of nanometers due to the fractional surface
coverage.\cite{quantum_island} The four samples, with the same thickness
but different lateral sizes, provide an excellent system to demonstrate
the suppression of spin relaxation process by lateral size constriction
down to 10s nm, without bringing significant spin scattering by the
edge defects as observed in narrow channel devices fabricated by etching
process.\cite{narrowing} 

\begin{figure}

\includegraphics[width=8cm]{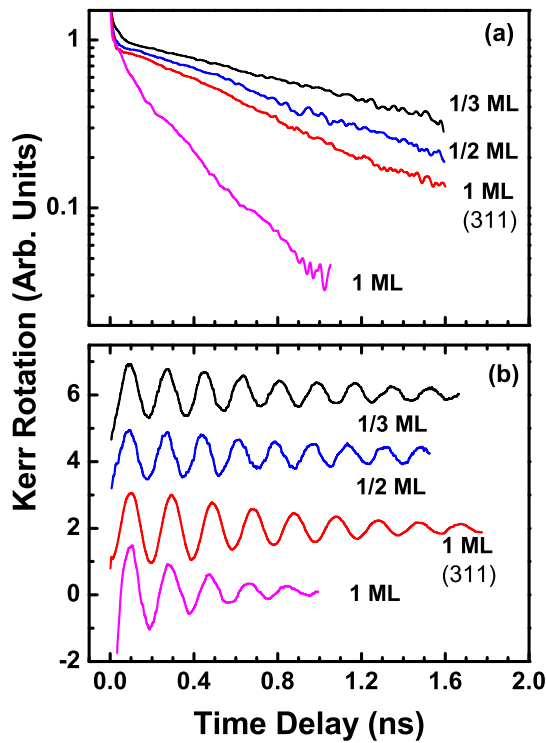}

\caption{(color online) Kerr rotation at (a) $B$=0$T$ and (b) $B$=0.82$T$
for samples with InAs thickness of 1/3 ML, 1/2 ML and 1ML, respectively.
The temperature is at 77$K$. The data of the sample on (311) substrate
is also shown. The pumping density is about $5\times10^{\text{16}}/cm^{\text{3}}$. }

\end{figure}

To study the coherent spin dynamics, an optical pump-probe spectroscopy
technique called time-resolved Kerr rotation (KR) spectroscopy is
used. A circularly polarized pump pulse generates a well-defined carrier
spin population, and the KR angle of a linearly polarized probe light
is detected by a balanced optical bridge. The time delayed probe pulse
reflects the time evolution of the projection of the electron spin
states, which precesses perpendicularly to the transverse external
magnetic field. Both pump and probe pulses are obtained from a tunable
mode-locked Ti:Sapphire laser with a pulse width about 150\textit{fs}
and a repetition rate of 80MHz. The pump beam size is about 30\textgreek{m}m
in focus and the size of the probe beam is tuned to be a little bit
smaller than that. The typical excitation powers are 0.5$\sim$10\textit{mW}
for the pump and 0.5\textit{mW} for the probe beams. To get an excellent
signal-to-noise ratio, a double lock-in detection technique is employed
with the amplitude modulation of the probe beam at 115\textit{Hz}
with an optical chopper and the polarization modulation of the pump
beam at 50\textit{KHz} with a photoelastic modulator. The measurements
were carried out in a magneto-optical cryostat in Voigt geometry with
a tunable photon energy of 1.45$\sim$1. 5\textit{eV}.

\section{results and discussion}

Figure 1(a) shows the KR signals at zero external field for samples
with InAs thickness of 1/3 ML, 1/2 ML and 1ML, respectively. The temperature
was 77\textit{K} and the pumping density was kept at low level with
pumping/probe ratio of 1:1 and the carrier density was estimated to
be about 5x10$^{\text{16}}$/cm$^{\text{3}}$. These data depict that
there is a fast decay of the spin polarization during the very first
10 picoseconds after excitation, and followed by a long-lifetime simple
exponential decay process. The fast decay originates from the loss
of hole spin polarization since the hole spin lifetime is short due
to strong valence band mixing and \textit{k}-dependent spin splitting.\cite{hole_spin}
The evolution of the KR angle thereafter can be described by a single
exponential decay $\theta_{k}=A_{0}\exp\left(-\nicefrac{t}{\tau_{s}}\right)$
for all the samples, where $A_{0}$ is proportional to the initial
amplitude of the electron-spin polarization, \textit{t} is the time
delay between the circularly polarized pump and the linearly polarized
probe pulse and $T_{2}^{*}$ is the electron spin relaxation time.
The independent evolution of the electron and hole spin polarization
indicates that we do not need to take the exciton spin as a constituent
as previously studied using time-resolved photoluminescence at these
temperatures.\cite{exciton} The spin relaxation time for the 1/3,
1/2 and 1ML samples on (001) surface are extracted to be 1500\textit{ps},
984\textit{ps} and 380\textit{ps}, respectively. The spin relaxation
time for the 1ML sample on (311)A surface is determined to be 860\textit{ps}.
These data clearly evidence that the electron spin lifetime is significantly
increased as the lateral size is gradually reduced, of which the mechanism
will be discussed later. 

Figure 1(b) exhibits the KR data at in-plane external magnetic field
$B$=0.82\textit{T} for the four samples. It shows a clear spin oscillation
under the transverse external magnetic field, which can be well described
by $S_{0}\exp\left(-t/T_{2}^{*}\right)\cos\left(g^{*}\mu_{B}Bt/\hbar\right)$,
where $S_{0}$ is the initial amplitude, $T_{2}^{*}$ is the inhomogeneous
transverse electron spin lifetime, $g^{*}$ stands for the effective
electron $g$-factor, $\mu_{B}$ and $\hbar$ are the Bohr magneton
and Planck constant, respectively. $T_{2}^{*}$s extracted from the
spin oscillation at 0.82\textit{T} are found to be almost the same
as the spin relaxation times measured at zero field. The magnitude
of the extracted electron $g^{*}$-factor for the four samples falls
in the narrow range of 0.46 to 0.48 which is remarkably different
from that of bulk InAs of about 15. The discrepancy of the electron
$g^{*}$ factor in these ultra-thin quantum wells from the bulk value
is attributed to the penetration of carrier wave function into the
barriers, quantum confinement energy,\cite{confinement} and strain\cite{strain15}
etc. 

The lack of inversion symmetry in such III-V compounds of zinc-blende
structures as GaAs and InAs results in spin splitting of the conduction
band via spin-orbit coupling. Spin-orbit coupling contributes a momentum
dependent effective magnetic field. This is the driving force for
spin relaxation in DP theory. The mechanism of the electron spin decoherence
occurs via the spin precession of carriers with finite crystal momentum
\textit{k} caused by the effective \textit{k}-dependent magnetic field
in an inversion-asymmetric material. Since spin polarization changes
during precession between scatterings, scattering acts against spin
relaxation and accordingly the spin lifetime is inversely proportional
to the momentum scattering time $\tau_{p}$(namely $\tau_{s}\sim\tau_{p}^{-1}$).
Although strictly speaking that \textit{k} is not a good quantum number
for submonolayer samples as a result of broken translational symmetry,
the mechanism is still good to describe the phenomena observed here.
A signature of this mechanism is that in the \textquoteleft{}\textquoteleft{}motional
narrowing\textquoteright{}\textquoteright{} regime where spin coherence
time greatly exceeds orbital scattering time $\tau_{p}$, cleaner
samples are expected to have shorter spin coherence time.\cite{momentumscattering}
It is naturally expected that scattering by boundaries and deformation
potentials will decrease $\tau_{p}$ in InAs submonolayer samples,
where the submonolayer exists in form of disks with lateral size of
tens of nanometer and consequently boundaries and deformation potentials
are enhanced with the decreased coverage. On the other hand, lateral
quantum confinement rising from the reduced lateral dimensions leads
to discrete momenta and energy levels, and therefore inhomogous broadending
of the electronic states which is an effective decay channel\cite{inhomogenous broading}
in 2D system, is gradually suppressed with decreased coverage. In
our experiment, the electron spin lifetime was found to be significantly
increased with the decreased InAs submomolayer coverage which leads
to a gradually reduced lateral size of the 2D system. Data presented
in Figure 1 supports the argument that a dimensionally constrained
DP mechanism slows down the spin relaxation in the motional narrowing
regime. 

\begin{figure}

\includegraphics[width=8.5cm]{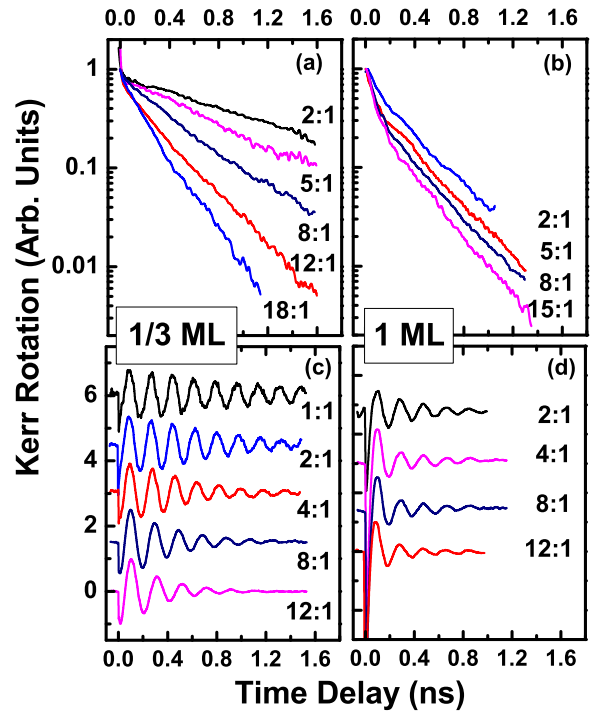}

\caption{(color online) Kerr rotation at $B=0T$ for (a) 1/3 ML and (b) 1ML
InAs samples, respectively, with various pumping density as indicated.
(c) and (d) are the pumping density dependent Kerr rotation of 1/3
ML and 1 ML InAs samples at $B=0.82T$\textit{,} respectively. The
temperature is 77 K.}

\end{figure}

Another important feature we found is that the spin lifetime in the
submonolayer InAs samples is strongly dependent on carrier density.
The KR data for 1/3 ML and 1ML InAs (001) samples at $B$=0\textit{T}
and 0.82\textit{T} are shown in Figures 2(a), 2(b), 2(c) and 2(d)
respectively, with various pumping density as indicated. In both zero
field and transverse external magnetic field cases, one can clearly
see that the higher pumping intensity, the faster decay of spin polarization
in the 1/3 ML sample. In contrast, the decay of the spin polarization
in the 1ML sample is not that sensitive to the pumping density. 

\begin{figure}

\includegraphics[width=8.5cm]{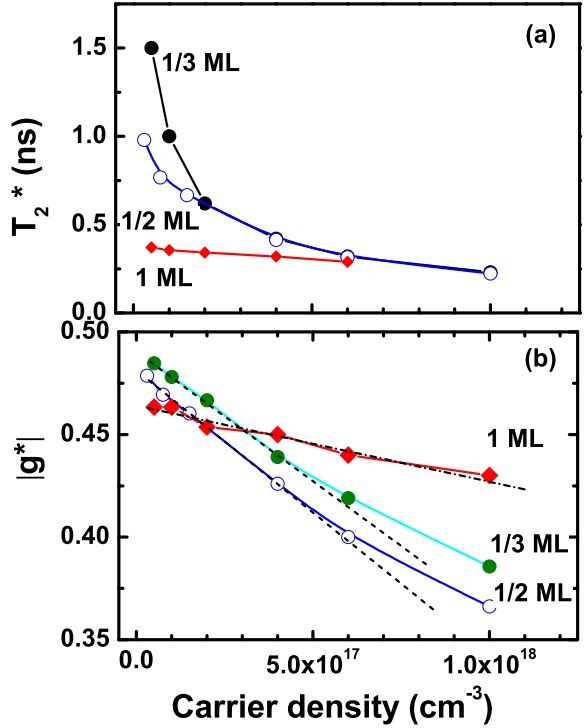}

\caption{(color online) (a) the derived electron spin decoherence time $T_{2}^{*}$
and (b) effective $g$-factor for the three samples on (100) substrates
with various carrier density. The temperature is 77$K$.}

\end{figure}

To get more insights into the spin relaxation mechanisms, we plot
the derived electron spin decoherence time $T_{2}^{*}$ and the effective
$g^{*}$-factor as a function of carrier density for the three samples
on (001) substrates. In Figure 3(a), we can find that $T_{2}^{*}$
in both 1/3 and 1/2 ML samples drop quickly as carrier density increases,
especially in the low carrier density region; while that of the 1
ML sample displays weak dependence on carrier density. We have shown
that the spin relaxation by DP mechanism in the submonolayers is suppressed
due to the lateral size confinement. The BAP mechanism involving electron-hole
exchange interaction may, however, be enhanced or even become dominant
in these quantum disk structures because of the strong interaction
between the spacially confined photoexcited electrons and holes. The
fact that $T_{2}^{*}$decreases with increasing carrier density also
agrees well with the expectation of the BAP process. As predicted
by Ref\cite{WuMW}, however, it is questioned that the effect of the
BAP mechanism at low temperature and high electron density is far
exaggerated in the literature due to the neglect of the nonlinear
terms in the spin-flip electron-hole exchange scattering. 

In Figure 3(b), we plot the carrier density dependence of $g^{*}$
for three different samples. The measured $g^{*}$ through transient
Kerr rotation should correspond to that at the Fermi energy. It has
been found that within a small energy range, the $g^{*}$-factor can
be approximated by $g^{*}=g_{0}+\beta E$ , where \textgreek{b} denotes
a constant and \textit{E} denotes the energy.\cite{g_E} The electron
density of states (DOS) follows $D\left(E\right)=dn/dE$, where\textit{
n} is the electron density. Given a measurement of $g^{*}$ as a function
of electron density, the electron DOS at the Fermi energy can therefore
be determined by $D\left(E\right)=\beta\left(dg^{*}/dn\right)^{-1}$,
where \textit{n} is the electron density up to the Fermi energy. In
a two-dimensional electron gas (2DEG), a linear dependence of $g^{*}$
on the 2DEG density is expected, given that the DOS of the 2DEG is
a constant independent of energy.\cite{DOS} This expectation agrees
well with our experimental results as shown in Figure 3(b). As the
InAs submonolayer exists in form of disks with lateral size of tens
of nanometer, the lateral quantum confinement splits the bands and
further modifies the DOS of the 2DEGS. Therefore it produces a more
rapid change in the $g^{\star}$-factor with electron density. The
observed slight non-linearity of $g^{*}$at higher density in the
submonolayer samples further supports the scenario of lateral quantum
confinement. 

To carefully identify the spin decoherence mechanism in the submonolayer
samples, we have measured temperature dependent KR with various pumping
density. Figure 4(a) shows some of the KR data of the 1/3 ML InAs
sample at different temperatures measured at B = 0.82\textit{T} with
pumping density about 1.5 x 10$^{\text{17}}$/cm$^{\text{3}}$. The
temperature dependence of the extracted electron spin decoherence
time $T_{2}^{\star}$ and effective $g^{\star}$-factor are shown
in (b) and (c), respectively. The $T_{2}^{*}$ of 1/3 ML InAs sample
with various pumping density as specified is also shown for comparison. 

\begin{figure}

\includegraphics[width=8.5cm]{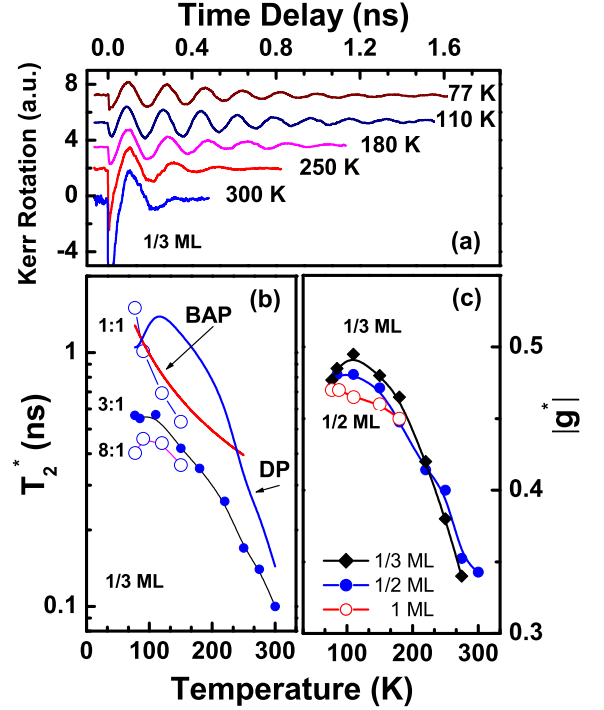}

\caption{(color online) (a) Kerr rotation of the 1/3 ML InAs sample at different
temperatures measured at B = 0.82\textit{T}. The pumping density is
about 1.5 $\times$10$^{\text{17}}$/cm$^{\text{3}}$. The temperature
dependence of the electron spin decoherence time $T_{2}^{*}$ and
the effective $g$-factor are shown in (b) and (c), respectively.
The $T_{2}^{*}$ of 1/3 ML InAs sample with various pumping density
at low temperature as specified is also shown for comparison. }

\end{figure}

BAP mechanism predicts spin relaxation time decreases rapidly with
increased temperature at low temperature, but is less sensitive to
temperature at higher temperature. The experimental data shown in
Figure 4(b) indicates that only the data at low temperature with low
pumping intensity (such as pump/probe=1:1) agrees with this dependence.
When the pumping intensity increases (pump/probe= 3:1), the spin lifetime
at low temperature shows flat temperature dependence. The spin lifetime
at higher carrier density (pump/probe=8:1) even displays a peak at
around 100 K. These observations clearly demonstrate that the BAP
mechanism is not the dominant process at higher carrier density or
at higher temperature. Actually, all the observations could be well
explained by Zhou and Wu\textquoteright{}s calculation\cite{WuMW}
including both BAP mechanism and DP mechanism using the kinetic spin
Bloch equations. The only discrepancy lies in that the BAP mechanism
plays a more important role in the submonolayer samples. The line
named BAP in Figure 4(b) is a fit of the data at low temperature and
low carrier density (pump/probe=1:1, pump$\sim0.5mW$), assuming $T^{-1}$
dependence of the BAP mechanism.\cite{assumption} 

We can find that the DP mechanism will determine the spin relaxation
process at high temperature. For the 1ML sample, DP mechanism is believed
to dominate at both low temperature and higher temperature. Figure
4(c) depicts the temperature dependence of the g$*$-factor for the
three InAs samples at pump/probe ratio of 3:1. The submonolayer samples
exhibit a hump-like dependence of the g$*$-factor on temperature;
while the 1ML sample shows a monotonic decrease at elevated temperatures. 

In conclusion, we have experimentally studied the spin relaxation
process in InAs ultrathin layer embedded in GaAs matrix. Long spin
relaxation time in the submonolayer structures has been observed and
been attributed to the suppression of DP process owing to constriction
of lateral dimensions. The electron spin relaxation time and effective
$g^{*}$-factor in submonolayer samples were found to be strongly
dependent on the photon-generated carrier density. The dependence
of spin relaxation on temperature has been examined and the related
mechanisms have been discussed. The clear coherent spin oscillation
at 300 K in the InAs ultrathin layer seems attractive for coherent
spin manipulation at room temperature. 
\begin{acknowledgments}
The authors thank M.W. Wu for helpful discussion. This work was supported
by Hong Kong GRF grant under HKU701308P,China NSF grants under 60706021
and 60876066. 
\end{acknowledgments}


\begin{thebibliography}{20}
\bibitem{assumption}This is only valid when $\tau_{p}$ is weak temperature
dependent. We think it is acceptable for a rough estimation since
$\tau_{p}$ in the submonolayer sample is mainly limited by the boundary
scattering.
\end{thebibliography}
\end{document}